\documentclass[a4paper,11pt]{article}
\usepackage{pos}
\def\beq{\begin{equation}}
\def\eeq{\end{equation}}
\def\beqa{\begin{eqnarray}}
\def\eeqa{\end{eqnarray}}

\title{Theoretical results for top-pair and top+$W$ production}

\author*{Nikolaos Kidonakis}

\affiliation{Department of Physics, Kennesaw State University,\\
Kennesaw, GA 30144, USA}

\emailAdd{nkidonak@kennesaw.edu}

\abstract{I present theoretical results for top-pair production as well as for the associated production of top quarks with $W$ bosons. Soft-gluon corrections from resummation are calculated through approximate N$^3$LO and added to fixed-order QCD results, and electroweak corrections are included at NLO. Top-quark transverse-momentum and rapidity distributions are also presented. In all cases the higher-order corrections are large, they reduce the scale dependence, and they improve agreement with recent data.}

\FullConference{42nd International Conference on High Energy Physics (ICHEP2024)\\
18-24 July 2024\\
Prague, Czech Republic\\}

\begin{document}

\maketitle

\section{Introduction}

The contributions of higher-order corrections to top-quark processes, such as $t{\bar t}$ production, $tW$ production, and $t{\bar t}W$ production are significant and improve the precision of theoretical predictions. Soft-gluon corrections are important for all top-quark processes that have been studied, and they approximate known exact results at NLO and NNLO very well. For a $2 \to n$ process with $p_a+p_b \to p_t+p_2 + \cdots +p_n$, where $p_t$ is the top-quark momentum, we define the threshold variable $s_4=(p_2 +\cdots +p_n+p_g)^2-(p_2 +\cdots +p_n)^2$ where an extra gluon with momentum $p_g$ is emitted.
At partonic threshold $p_g \to 0$ and, thus, $s_4 \to 0$. The soft corrections involve logarithms of the form $[\ln^k(s_4/m_t^2)/s_4]_+$ with $m_t$ the top-quark mass and $k \le 2n-1$ for the order $\alpha_s^n$ corrections. We resum these corrections in single-particle-inclusive kinematics for the double-differential cross section and use them to derive approximate NNLO (aNNLO) and approximate N$^3$LO (aN$^3$LO) predictions for total cross sections and top-quark differential distributions.

\section{Soft-gluon corrections}

The factorized cross section for a general top-quark production process can be written as
\beq
d\sigma_{pp \to tX}=\sum_{a,b} \; 
\int dx_a \, dx_b \,  \phi_{a/p}(x_a, \mu_F) \, \phi_{b/p}(x_b, \mu_F) \, 
d{\hat \sigma}_{ab \to tX}(s_4, \mu_F) 
\eeq
where the $\phi$'s are parton distribution functions (pdf), $\mu_F$ is the factorization scale, and ${\hat \sigma}$ is the partonic cross section.
We take Laplace transforms $d{\tilde{\hat \sigma}}_{ab \to tX}(N)=\int (ds_4/s) \; e^{-N s_4/s} d{\hat \sigma}_{ab \to tX}(s_4)$ 
and ${\tilde \phi}(N)=\int_0^1 e^{-N(1-x)} \, \phi(x) \, dx$ with transform variable $N$, 
and we write the parton-parton cross section as 
\beq
d{\tilde \sigma}_{ab \to tX}(N)= {\tilde \phi}_{a/a}(N_a, \mu_F) \; {\tilde \phi}_{b/b}(N_b, \mu_F) \; d{\tilde{\hat \sigma}}_{ab \to tX}(N, \mu_F) \, .
\label{ppfac}
\eeq
A further refactorization for the cross section, 
\beq 
d{\tilde\sigma}_{ab \to tX}(N)
= {\tilde \psi}_a(N_a,\mu_F) \, {\tilde \psi}_b(N_b,\mu_F) \; 
{\rm tr} \left\{H_{ab \to tX}\left(\alpha_s(\mu_R)\right) \, 
{\tilde S}_{ab \to tX}\left(\frac{{\sqrt s}}{N \mu_F}\right) \right\} \, , 
\label{pprefac}
\eeq
is given in terms of new functions $\psi$ for collinear emission from incoming partons, a hard function $H_{ab \to tX}$, and a soft function $S_{ab \to tX}$ for noncollinear soft gluons. The hard and soft functions are matrices in the space of color flow. In the case of collinear emission from any massless final-state particles we also need a function $J$, but this does not apply to the processes studied here.

From Eqs. (\ref{ppfac}) and (\ref{pprefac}) we, thus, have the formula
\beq
d{\tilde{\hat \sigma}}_{ab \to tX}(N,\mu_F)=
\frac{{\tilde \psi}_{a/a}(N_a, \mu_F) \, {\tilde \psi}_{b/b}(N_b, \mu_F)}{{\tilde \phi}_{a/a}(N_a, \mu_F) \, {\tilde \phi}_{b/b}(N_b, \mu_F)} \; {\rm tr} \left\{H_{ab \to tX}\left(\alpha_s(\mu_R)\right) \, 
{\tilde S}_{ab \to tX}\left(\frac{\sqrt{s}}{N \mu_F} \right)\right\} \, .
\eeq
$S_{ab \to tX}$ satisfies the renormalization-group equation, with $\mu_R$ the renormalization scale, 
\beq
\left(\mu_R \frac{\partial}{\partial \mu_R}
+\beta(g_s)\frac{\partial}{\partial g_s}\right)\,S_{ab \to tX}
=-\Gamma^{\dagger}_{\! S \, ab \to tX} \, S_{ab \to tX}-S_{ab \to tX} \, \Gamma_{\! S \, ab \to tX} \, .
\eeq
The soft anomalous dimension $\Gamma_{\! S \, ab \to tX}$ controls the evolution of the soft function via the above equation and, thus, gives the exponentiation of logarithms of $N$ \cite{NKGS,NK2ltt}.

Renormalization-group evolution leads to resummation, and we find the expression \cite{NKGS,NK2ltt}
\beqa
d\tilde{\hat \sigma}_{ab \to tX}^{\rm resum}(N) & = &
\exp\left[\sum_{i=a,b} E_{i}(N_i)\right] \, 
\exp\left[\sum_{i=a,b} 2 \int_{\mu_F}^{\sqrt{s}} \frac{d\mu}{\mu} \gamma_{i/i}(N_i)\right] \; \times {\rm tr} \left\{H_{ab \to tX} \left(\alpha_s(\sqrt{s})\right)\right.
\nonumber \\ && \hspace{-33mm} \left.
\times \exp \left[\int_{\sqrt{s}}^{{\sqrt{s}}/N} \! \frac{d\mu}{\mu} \Gamma_{\! S \, ab \to tX}^{\dagger} \left(\alpha_s(\mu)\right)\right] 
{\tilde S}_{ab \to tX} \left(\alpha_s\left(\frac{\sqrt{s}}{N}\right)\right)  \exp \left[\int_{\sqrt{s}}^{{\sqrt{s}}/N} \! \frac{d\mu}{\mu} \Gamma_{\! S \, ab \to tX} \left(\alpha_s(\mu)\right)\right] \right\}. 
\eeqa
The resummed cross section is then expanded at fixed-order through N$^3$LO and transformed back to physical space. Many top processes have been studied with calculations of corrections for total and differential cross sections. These include top-pair production: $t{\bar t}$ in the SM through aN$^3$LO, and $t{\bar t}$ in SMEFT at aNNLO; top-pair+$X$: $t{\bar t}\gamma$ at aNNLO and $t{\bar t}W$ at aN$^3$LO; single top: $t$- and $s$-channel and $tW$ through aN$^3$LO; single-top+$X$: $tqH$, $tq\gamma$, and $tqZ$ through aNNLO; single-top BSM: $t\gamma$, $tZ$, $tZ'$, $tg$ through aNNLO, and $tH^-$ through aN$^3$LO; see Ref. \cite{NKrev} for reviews of the theory of soft-gluon corrections that has been used for all these processes and many others as well.

\section{$t{\bar t}$ production}

\begin{figure}[htpb]
\begin{center}
\includegraphics[width=72mm]{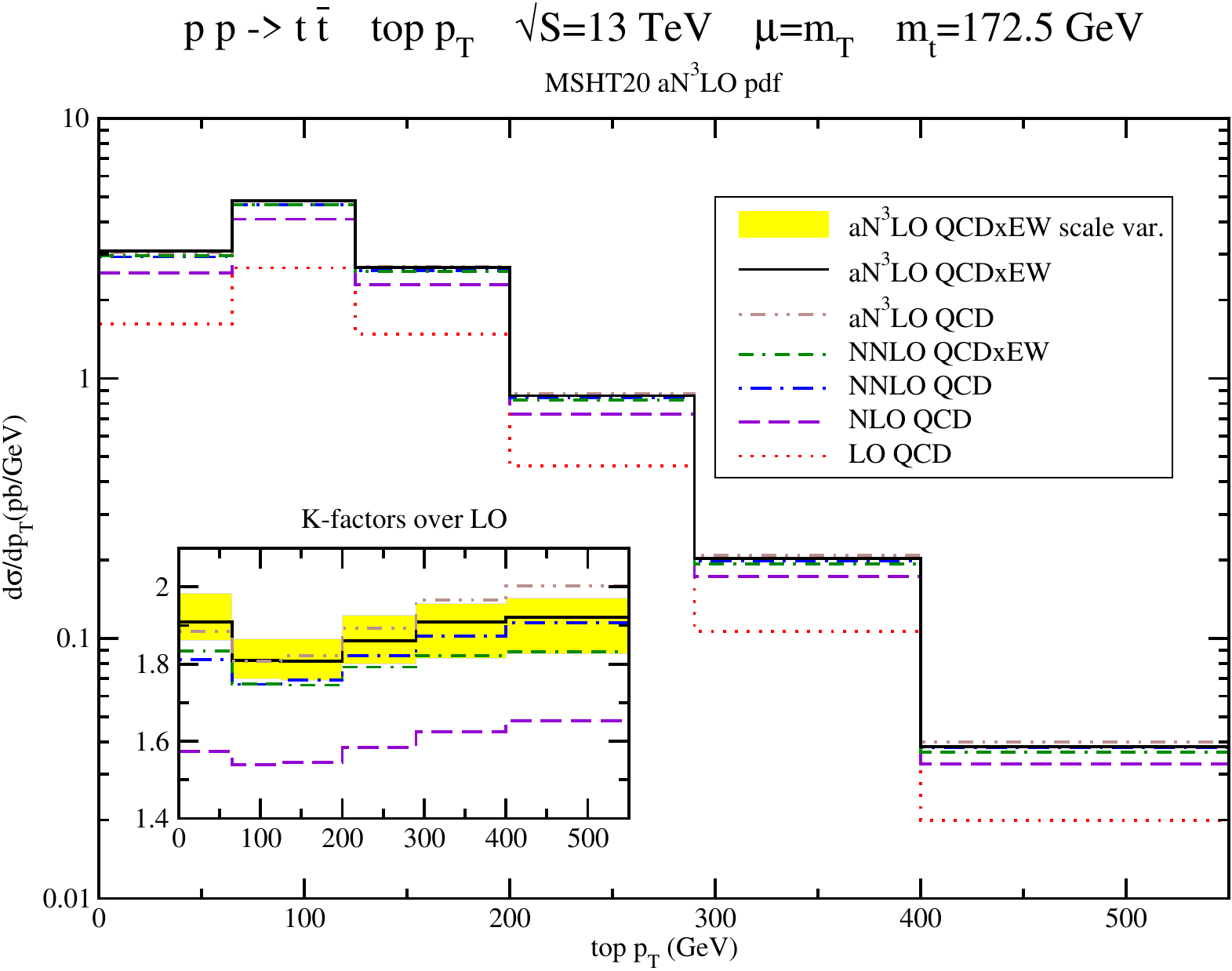}
\hspace{2mm}
\includegraphics[width=72mm]{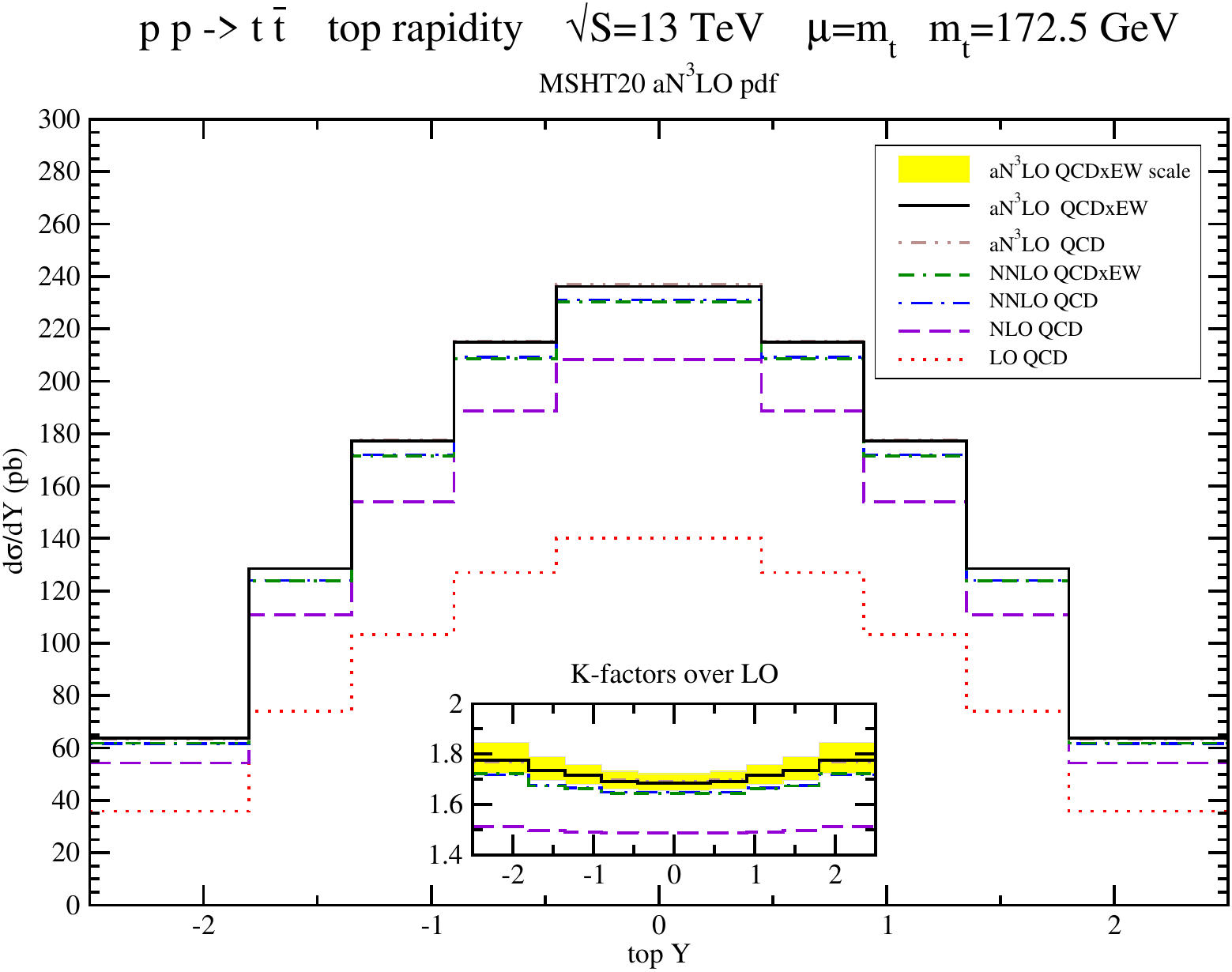}
\caption{Top-quark $p_T$ (left) and rapidity (right) distributions in top-pair production at 13 TeV LHC energy.}
\label{pTyttbar}
\end{center}
\end{figure}

Calculations of soft-gluon corrections for top-quark differential cross sections go back thirty years \cite{NKtoppty}. The soft anomalous dimension matrix is $2\times 2$ for the $q{\bar q} \to t{\bar t}$ channel and $3\times 3$ for the $gg \to t{\bar t}$ channel; they were calculated at one loop in the mid-90's \cite{NKGS} and at two loops fifteen years ago \cite{NK2ltt}. There are partial results at three loops \cite{NKrev}, and the four-loop massive cusp anomalous dimension (determined from its asymptotics in Ref. \cite{NK4lcusp}) contributes to the four-loop result.

The NLO expansions of the resummed cross section agree with exact NLO results very well, and the NNLO expansions provide aNNLO results that predicted the exact NNLO to very high accuracy (percent or per mille) for total cross sections and top-quark $p_T$ and rapidity distributions. By further adding the third-order soft-gluon corrections to the exact NNLO result, we obtain aN$^3$LO predictions \cite{NKaN3LO} which are the state of the art, and to which we also add electroweak corrections \cite{KGT}.

The aN$^3$LO QCD + NLO EW cross section with $\mu=m_t$ and scale and pdf uncertainties is \cite{KGT} with MSHT20 aN$^3$LO pdf \cite{MSHT20an3lo} at 13 TeV, $802^{+22}_{-17} {}^{+16}_{-17}$ pb, and at 13.6 TeV,  $886^{+24}_{-19} {}^{+18}_{-20}$ pb; with MSHT20 NNLO pdf \cite{MSHT20nnlo} at 13 TeV, $836^{+23}_{-18} {}^{+17}_{-11}$ pb, and at 13.6 TeV, $925^{+25}_{-20} {}^{+18}_{-12}$ pb; with CT18 NNLO pdf \cite{CT18nnlo} at 13 TeV, $842^{+23}_{-18} {}^{+18}_{-16}$ pb, and at 13.6 TeV, $932^{+25}_{-20} {}^{+20}_{-18}$ pb; and with NNPDF4.0 NNLO pdf \cite{NNPDF40nnlo} at 13 TeV, $816^{+23}_{-18} {}^{+5}_{-4}$ pb, and at 13.6 TeV, $904^{+25}_{-20} {}^{+5}_{-5}$ pb. Furthermore, with PDF4LHC21 NNLO pdf \cite{PDF4LHC21nnlo} the result at 13 TeV is  $837^{+23}_{-18} {}^{+20}_{-16}$ pb, and at 13.6 TeV it is  $926^{+25}_{-20} {}^{+22}_{-17}$ pb.  

Figure \ref{pTyttbar} displays theoretical results through aN$^3$LO QCD $\times$ EW for the top-quark $p_T$ and rapidity distributions in $t{\bar t}$ production \cite{KGT} at 13 TeV LHC energy using MSHT20 aN$^3$LO pdf \cite{MSHT20an3lo}.

\section{$tW$ production}

Calculations of soft-gluon corrections for $tW$ production have been available for close to two decades \cite{NKtW,NKNY}, and the soft anomalous dimension for $tW$ production is known to three loops \cite{NKtW}. 

\begin{figure}[htpb]
\begin{center}
\includegraphics[width=72mm]{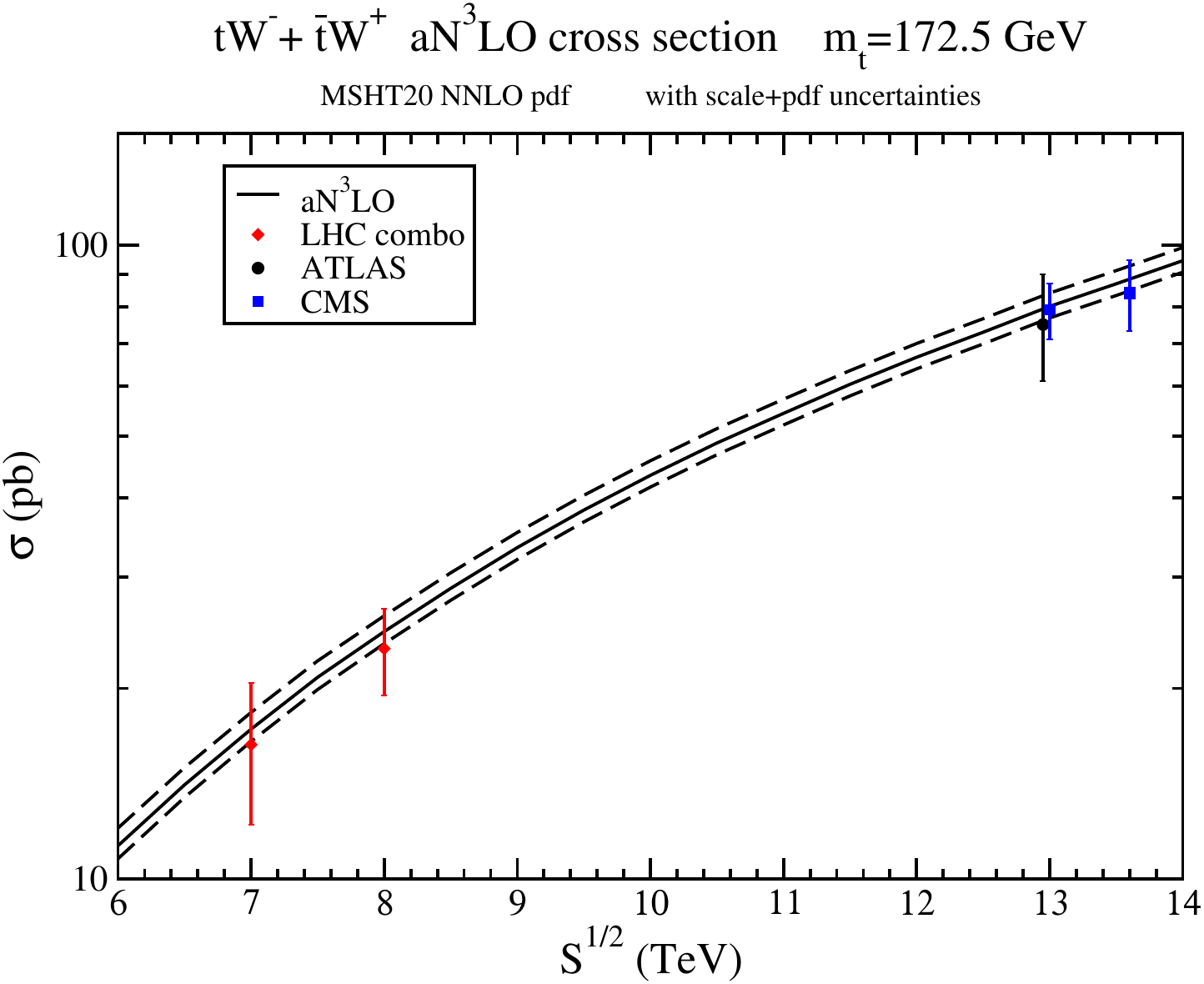}
\caption{$tW^-+{\bar t}W^+$ cross section at aN$^3$LO with scale and pdf uncertainties at LHC energies.}
\label{tW}
\end{center}
\end{figure}

Figure \ref{tW} shows the theoretical $tW^-+{\bar t}W^+$ production cross sections at aN$^3$LO \cite{NKNY} at LHC energies and compares them with LHC data at 7, 8, 13, and 13.6 TeV \cite{ATLASCMS78tW,ATLAS13tW,CMS13136tW}.

The aN$^3$LO cross section for $tW^-+{\bar t}W^+$ production with $\mu=m_t$ and scale and pdf uncertainties 
with MSHT20 NNLO pdf \cite{MSHT20nnlo} at 13 TeV is $79.5^{+1.9}_{-1.8} {}^{+2.0}_{-1.4}$ pb, and at 13.6 TeV it is $87.6^{+2.0}_{-1.9} {}^{+2.1}_{-1.5}$ pb; with MSHT20 aN$^3$LO pdf \cite{MSHT20an3lo} at 13 TeV it is $77.3^{+1.9}_{-1.8} {}^{+2.0}_{-2.1}$ pb, and at 13.6 TeV it is $85.6^{+2.0}_{-1.9} {}^{+2.2}_{-2.3}$ pb; and with PDF4LHC21 pdf \cite{PDF4LHC21nnlo} at 13 TeV it is $79.3^{+1.9}_{-1.8} {}^{+2.2}_{-2.2}$ pb, and at 13.6 TeV it is $87.9^{+2.0}_{-1.9} {}^{+2.4}_{-2.4}$ pb.  

\section{$t{\bar t}W$ production}

The observation of $t{\bar t}W$ events at the LHC has shown that the measurements are significantly higher than theoretical predictions at NLO. 
The QCD corrections at NLO are large while the electroweak corrections are smaller but significant.
Further improvement in theoretical accuracy can be obtained by the inclusion of higher-order soft-gluon corrections \cite{NKCF} using the formalism for 2 $\to$ 3 processes \cite{MFNK2n}. We calculate aN$^3$LO QCD corrections and further add to them NLO electroweak corrections to provide state-of-the-art theoretical predictions.

\begin{figure}[htpb]
\begin{center}
\includegraphics[width=72mm]{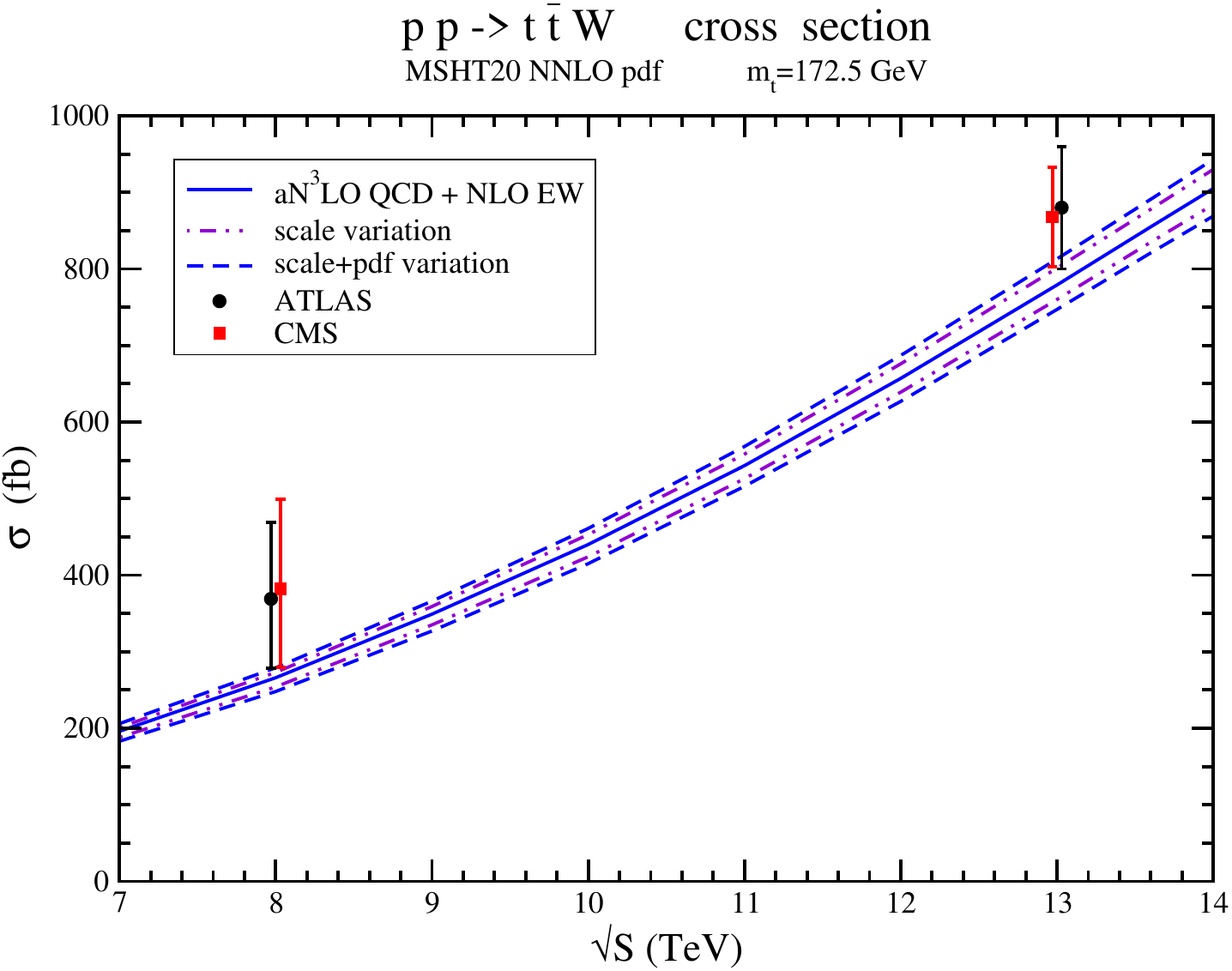}
\hspace{2mm}
\includegraphics[width=72mm]{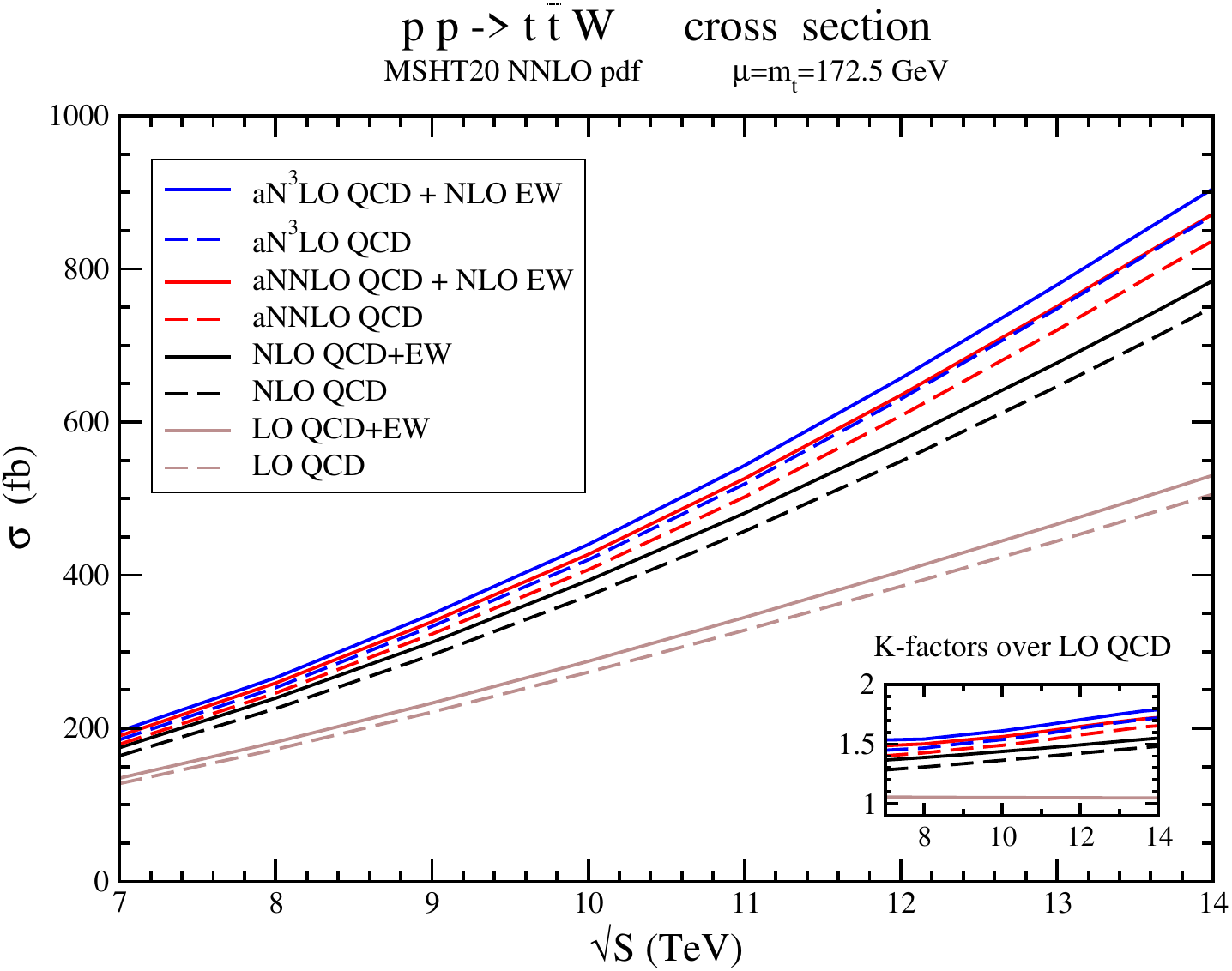}
\caption{Cross sections through aN$^3$LO QCD+NLO EW for $t{\bar t}W$ production at LHC energies.}
\label{ttbarW}
\end{center}
\end{figure}

Figure \ref{ttbarW} shows the theoretical predictions for the total $t{\bar t}W$ cross section using MSHT20 NNLO pdf \cite{MSHT20nnlo} and a comparison with data at 8 and 13 TeV from ATLAS \cite{ATLASttW} and CMS \cite{CMSttW}. We note the large $K$-factors and the improved agreement with data at aN$^3$LO. 

At 13.6 TeV with $\mu=m_t$, the NLO QCD corrections increase the LO result by 47\%, the aNNLO QCD corrections by a further 17\% (and consistent with the partial NNLO result in \cite{BDGK}), and the aN$^3$LO QCD corrections by an extra 6\%, while the electroweak NLO corrections provide 7\%. Thus, the total aN$^3$LO QCD+NLO EW cross section is 78\% bigger than LO QCD.

In the comparison with 8 and 13 TeV CMS and ATLAS data, the NLO and even aNNLO results are not sufficient; we need aN$^3$LO corrections to describe the data. At 8 TeV, the measured cross section from CMS is $382^{+117}_{-102}$ fb and from ATLAS it is $369^{+100}_{-91}$ fb. The theoretical prediction at aN$^3$LO QCD + NLO EW with central scale $\mu=m_t$ is $266^{+7}_{-12}{}^{+6}_{-6}$ fb.

At 13 TeV, CMS finds $868 \pm 65$ fb, with $553 \pm 42$ fb for $t{\bar t}W^+$ and $343 \pm 36$ fb for $t{\bar t} W^-$, 
while ATLAS finds $880 \pm 80$ fb, with $583 \pm 58$ fb for $t{\bar t}W^+$ and $296 \pm 40$ fb for $t{\bar t} W^-$. 
The theoretical prediction at aN$^3$LO QCD + NLO EW with central scale $\mu=m_t$ is $779^{+22}_{-19}{}^{+12}_{-13}$ fb, with $517^{+14}_{-12}{}^{+8}_{-9}$ fb for $t{\bar t}W^+$
and $262^{+8}_{-7}{}^{+4}_{-4}$ fb for $t{\bar t} W^-$.

Top-quark $p_T$ and rapidity distributions in $t{\bar t}W$ production at 13 and 13.6 TeV have also been calculated in \cite{NKCF}; the $K$-factors decrease at larger top $p_T$ while they increase at larger rapidities.

Finally, we note that soft-gluon corrections are important for the related processes of $t{\bar t}Z$ production.

\acknowledgments
This material is based upon work supported by the National Science Foundation under Grant Nos. PHY 2112025 and PHY 2412071.

\end{document}